\documentstyle[aps,multicol,epsf]{revtex}
\draft
\def\beginwide{
        \end{multicols} \vspace*{-0.5cm} \noindent
        \rule{3.5in}{.1mm}\rule{.1mm}{5mm} \widetext \medskip }
\def\beginwidetop{
        \end{multicols} \vspace*{-0.5cm} \noindent
        \widetext \medskip }
\def\endwide{
        \hspace*{3.35in}~\rule[-5mm]{.1mm}{5mm}\rule{3.5in}{.1mm}
        \begin{multicols}{2} \vspace*{-1.0cm} \noindent }
\def\endwidebottom{
        \begin{multicols}{2} \vspace*{-1.0cm} \noindent }

\newcommand{\beq}{\begin{equation}}
\newcommand{\eeq}{\end{equation}}
\newcommand{\bdis}{\begin{displaymath}}
\newcommand{\edis}{\end{displaymath}}
\newcommand{\bea}{\begin{eqnarray}}
\newcommand{\eea}{\end{eqnarray}}
\newcommand{\barr}{\begin{array}}
\newcommand{\earr}{\end{array}}
\begin{document}

\title{Universality in sandpiles}

\author{Alessandro Chessa$^{(1)}$, H. Eugene Stanley$^{(2)}$, 
Alessandro Vespignani$^{(3)}$, and Stefano Zapperi$^{(4)}$}

\address{
1)Dipartimento di Fisica and Unit\'a INFM, Universit\'a di Cagliari, 
Via Ospedale 72, 09124 Cagliari, Italy\\
2)Center for Polymer Studies and Department of Physics
Boston University, Boston, MA 02215\\
3)The Abdus Salam International Centre for Theoretical Physics (ICTP), 
P.O. Box 586, 34100 Trieste, Italy\\
4)PMMH-ESPCI, 10 Rue Vauquelin, 75234 Paris CEDEX 05, France.
}

\date{\today}

\maketitle
\begin{abstract}
We perform extensive numerical simulations 
of different versions of the sandpile model. We find that previous
claims about universality classes are unfounded, since the method
previously employed to analyze the data 
suffered a systematic bias. We identify the correct 
scaling behavior and conclude that sandpiles with stochastic  and 
deterministic toppling rules belong  to the same universality class.
\end{abstract}


%
%
%

\begin{multicols}{2}

Sandpile automata \cite{btw} are among the simplest models 
to describe avalanche propagation, a phenomenon of upsurging
experimental interest in a wide range of fields \cite{exp}. In the stationary
state, after suitable tuning of the driving fields \cite{vz}, 
these models display 
critical behavior in the avalanche statistics. As for ordinary
critical phenomena, it is possible to define a set of scaling
exponents that characterize that large scale behavior of the system
\cite{vz}.

The precise identification of universality classes in sandpile models 
\cite{btw} is an unresolved issue. From a theoretical standpoint,
it would be unusual that small modifications in the dynamical
rules of the model could lead to different universality classes.
Real-space renormalization group calculations \cite{vzp} suggest that
different sandpile models, such as the Bak, Tang and 
Wiesenfeld (BTW) \cite{btw} and the Manna
\cite{manna} models, all belong to the same universality class.
This result is also confirmed by a recently-proposed 
field theory approach \cite{ron}, 
which shows that all sandpile models
\cite{nota_dir} are described by the same effective field theory 
at the coarse grained level. Universality is also found between
BTW (discrete) and Zhang \cite{zhang} (continuous) models in the 
dynamical renormalization group calculations of Ref.~\cite{diaz}.

The results obtained by numerical simulations are unclear.
Early large scale numerical simulations
of the Manna \cite{manna} and BTW models \cite{grasma}, show that
the avalanche distributions are described by the same exponents
for the power law decay and the scaling of the cutoffs.
These results were questioned by Ben Hur and Biham \cite{ben} 
who analyzed the scaling of conditional expectation values \cite{cfj}
of various quantities. They found significant differences in the
exponents for the two models and therefore proposed a new classification
of universality in sandpile models in which models with 
stochastic update rules, such as the Manna model,
fall in a universality class different from that of Abelian
models, such as the BTW \cite{dhar}. The method was later applied to
the Zhang model that was declared ``non-universal'' \cite{biham}.
This results pose a puzzling problem, since they contradict 
all the existing theories and do not agree with the scaling predicted 
analyzing avalanche distributions \cite{manna,grasma}.

Here we present large scale numerical simulations of the BTW and
Manna sandpile models, with the goal of settling the issue of universality.
First we show that the method of conditional expectation values,
introduced in Ref.~\cite{cfj} and used in Ref.~\cite{ben},
is systematically biased by non-universal corrections and 
does not provide indications on universality classes. By removing
the bias, we provide evidence that the BTW and Manna models are universal.
We confirm this conclusion by data collapse and  
moment analysis of the distributions \cite{att}.

Sandpile models are defined on $d-$dimensional hypercubic lattice.
On each site $i$ of the lattice we define an integer variable 
$z_i$ which we call ``energy''. 
At each time step an energy grain is added on a randomly chosen site
($z_i\to z_i+1$). When one of the sites reaches or exceeds a threshold
$z_c$ a ``toppling'' occurs: $z_i=z_i-z_c$ and $z_j=z_j+1$, where 
$j$ represents the nearest neighbor sites of site $i$. In the BTW model
$z_c=2d$ and each nearest neighbor receives a grain after the toppling
of the site $i$. In the Manna model $z_c=2$ and therefore only two
randomly chosen neighboring sites receive a grain. A toppling
can induce nearest-neighbor sites to topple on
their turn and so on, until all the lattice 
sites are below the critical threshold.
This process is called an avalanche. A slow driving is usually imposed, 
so that grains are added only when all the sites are below the threshold.
The model is conservative and energy is dissipated only at boundary
sites \cite{btw}. 
Here, we perform numerical simulations of 
two-dimensional  Manna and BTW models with open boundary conditions 
and conservative dynamics. The lattice size ranges 
from $L=128$ to $L=2048$ in both models.
In each case, statistical distributions are obtained averaging over 
$10^7$ nonzero avalanches. 

Avalanches in sandpile models are usually characterized by three
variables: the number of topplings $s$, the area $a$ affected
by the avalanche, and the avalanche duration $T$. 
The probability distribution of each of
these variables is usually described as a power law with a cutoff
\beq
P(x)=x^{-\tau_x} {\cal G}(x/x_c),
\label{pl}
\eeq
where $x=s,a,T$. When
the system size $L$ goes to infinity 
the cutoff $x_c$ diverges as $x_c\sim L^{\beta_x}$. 
Under the finite size scaling (FSS) assumption of Eq.~(\ref{pl}), the 
set of exponents $\{\tau_x,\beta_x\}$ defines the universality
class of the model. 

In two dimensions,
an accurate numerical 
determination of the power law exponents in Eq.(\ref{pl})
proved to be a difficult task \cite{manna,grasma,cmvz,lubeck}, due
to the large deviations at the lower and upper cutoffs. For this
reason,  Christensen et al. \cite{cfj} in order to distinguish among 
universality classes proposed
a more refined numerical analysis based on the evaluation 
of the expectation value $E(x|y)$ of the variable $x$ restricted to 
all the avalanches with variable $Y=y$, 
where $\{X,Y\}=\{s,a,T\}$ \cite{cfj}. 
It is assumed that $E(x|y)\sim y^{\gamma_{xy}}$ and the exponents
$\gamma_{xy}$ are used to distinguish among universality classes \cite{ben}.
These exponents satisfy the scaling relations
$\gamma_{xy}=\gamma_{yx}^{-1}$ and $\gamma_{xz}=\gamma_{xy}\gamma_{yz}$.

If the conditional probability distribution $p(x|y)$ is sufficiently 
peaked, then $\gamma_{xy}$ is well-defined 
and to each value of the variable $x$ we can unambiguously
associate a value of the  variable $y$ (i.e. $x \sim y^{\gamma_{xy}}$). 
In particular, the cutoff of the distributions should be related
by the same exponents (i.e.  $x_c \sim y_c^{\gamma_{xy}}$), 
which implies $\gamma_{xy} = \beta_x/\beta_y$. 
For instance, we have $\gamma_{sa} = \beta_s/2$, since 
in two dimensions avalanches are compact for both the BTW \cite{grasma} 
and Manna model \cite{ben}, so that $\beta_a=2$.
The data collapse analysis shows the BTW and Manna model
both share the same exponent $\beta_s\simeq 2.7$ \cite{manna,grasma,cmvz}
which implies $\gamma_{sa} \simeq 1.35$. 
On the contrary, Refs.~\cite{ben,biham} found $\gamma_{sa}\simeq 1.06$ 
for the BTW model and $\gamma_{sa}=1.24$ for
the Manna model, which would yield two different universality classes
for the two models. Less marked differences 
were also observed for the other exponents 
$\gamma_{xy}$ \cite{ben,biham}.

In order to resolve this paradox, we return to the hypothesis
underlying the use of conditional expectation values:
$p(x|y)$ must be symmetric and strongly peaked around the average value.
We checked numerically that this assumption is {\it not} fulfilled:
in the BTW model the distribution $p(s|a)$ 
is maximum for $s=a$ and decreases for
$s>a$, with a characteristic value $s^*$ scaling as $a^{\beta_s/2}$ 
(see Fig.~1). 
The distribution is not symmetric (see also Ref.~\cite{lubeck}), 
consistent with the constraint
$s \geq a$ (the avalanche area can not be greater than the number of
topplings). Similar considerations apply as well to other quantities
(i.e. $a\geq T$, $s \geq T$) whose conditional probability
distributions show asymmetry although less marked.

To understand the effect of non-symmetric distributions on
conditional expectation values, consider a distribution
of the form 
\beq
p(x|y)= \theta(x-y)f((x-y)/x^*)/x^* 
\label{pxy}
\eeq 
where the characteristic value
scales as $x^*(y) \sim y^{\gamma_{xy}}$ and $\theta(x)$
is the step function . The factor $1/x^*$ ensures normalization for any $y$
\beq  
\int dx p(x|y) = 1,
\eeq
so that the conditional expectation value is given by
\beq
E(x|y)\equiv\int_y^\infty dx \frac{x}{x^*} f((x-y)/x^*).
\eeq
Performing the substitution $z=x-y$, we obtain
\beq
E(x|y)=y+\int_0^\infty dz \frac{z}{x^*} p(z/x^*)=y+Cy^{\gamma_{xy}},
\eeq
where $C$ is a non-universal constant.
 
In the BTW model $p(s|a)$ has the form of  Eq.~(\ref{pxy}) as  
is shown by performing data collapse analysis (see inset of Fig.~1). 
Thus, we can easily subtract the linear bias from the expectation
value in order to obtain the correct scaling behavior
to be compared with that of the Manna model (Fig.~2), 
whose conditional distributions appear to be symmetric.
Data from avalanche areas up to $a\simeq 10^6$ provide striking evidence 
that both models share the same asymptotic behavior 
with an exponent $\gamma_{sa}=1.35\pm0.05$,
in agreement with other published results \cite{manna,grasma,cmvz,lubeck}.
The scaling of the other expectation values is also biased as it is 
apparent from the bending in the curves reported in 
Refs.\cite{ben,biham}. The correction of the bias is not so
straightforward as in the case we have discussed but can be obtained
from the analysis of $p(x|y)$. This discussion clearly shows that 
conditional expectation values are not a reliable method 
to determine the universality class of a model. 

To confirm the conclusion about a single universality class 
we perform the moment analysis on the distributions
$P(x,L)$ in close analogy with the recent work of 
De Menech et al.\cite {att} on the two-dimensional BTW model. 
Here we apply the moments analysis on both BTW and Manna models, taking 
advantage of the large sizes reached in our numerical simulations.
We define the $q$-moment of $x$ on a lattice of size $L$ as
$\langle x^q\rangle_L=\int x^q P(x)dx$. 
If FSS hypothesis (Eq.~(\ref{pl})) is valid, at 
least in the asymptotic limit ($x\to\infty$), we can transform  
$z=x/L^{\beta_x}$ and obtain
\begin{equation}
\langle x^q\rangle_L= L^{\beta_x(q+1-\tau)}\int z^{q+\tau}{\cal G}(z)dz
\sim L^{\beta_x(q+1-\tau)},
\label{eq:xq}
\end{equation}
or in general $\langle x^q\rangle_L\sim L^{\sigma_x(q)}$.
The exponents $\sigma_x(q)$ can be obtained as the slope of the 
log-log plot of $<x^q>_L$ versus $L$. Using  Eq.~(\ref{eq:xq}), 
we obtain $\langle x^{q+1}\rangle_L/\langle x^q\rangle_L
\sim L^{\beta_x}$ or $\sigma_x(q+1)-\sigma_x(q)=\beta_x$,
so that the slope of $\sigma_x(q)$ as a function of $q$
is the cutoff exponent
$\beta_x =\partial\sigma_x(q)/\partial q$.
This is in general not true for small $q$ because the integral in
Eq.~(\ref{eq:xq}) is dominated by the lower cutoff. 
In particular, corrections to scaling  of the type 
$\langle x^q\rangle_L\sim L^{\sigma_x(q)}F(L)$
are important for $q \leq\tau_x-1$.
For instance, when  $q\simeq \tau_x-1$, logarithmic corrections give rise 
to effective exponents up to very large  lattice sizes. 
Finally, normalization
imposes $\sigma_x(0)=0$.

In Fig.~3 we show the results obtained from the moment analysis 
of the distribution $P(s)$ for the  Manna and the BTW model.
In this case we can use the exact result 
$\langle s\rangle\sim L^2$, which implies $\sigma_s(1)=2$, 
as a test for the convergence of our simulations
to the asymptotic scaling regime. This relation is
fulfilled and the $\sigma_s(q)$ of the two models are indistinguishable 
for $q\geq 1$, indicating universal scaling behavior. 
We observe small deviations for small $q$ which are 
due to the non-universal lower cutoff.
By measuring slope of $\sigma_s(q)$, we obtain 
$\beta_s\simeq 2.7$. This value is larger than the value reported in 
Ref.~\cite{att} (i.e. $D\simeq 2.5$), 
where small lattice sizes have been used. 
We have repeated the same analysis for the $P(T,L)$ and the $P(a,L)$
and the measured cutoff exponents $\beta_t$ and $\beta_a$
are reported in Table I. Also in this case the exponents for the two models 
share the same values within error bars, confirming the presence of a single 
universality class.

As a final consistency test, we used the data collapse method in order 
the check the FSS hypothesis, 
which states that rescaling  $q_x\equiv x/L^{\beta_x}$ 
and $P_{q_x}\equiv P(x,L)L^{\beta_x\tau_x}$, the data for different $L$
must collapse onto universal curves.
If FSS is verified, we can compute the exponent $\tau_x$ from the scaling 
relation $(2-\tau_x)\beta_x=\sigma_x(1)$, that should be satisfied for 
enough large sizes. Using the values of $\beta_x$ reported in Table I and 
the values obtained for $\sigma_x(1)$ we  
find the exponents $\tau_x$ to be inserted in the data collapse.
For instance, using the exact result  $\sigma_s(1)=2$ and the
estimated $\beta_s=2.7$, we obtain $\tau_s=1.27$. The data collapse
with these values is satisfactory for both models (see Fig.4).   

In the same way, we obtain very good data collapse for the Manna 
model $P(a)$ and $P(t)$ distributions, yielding $\tau_t=1.5$ and 
$\tau_a=1.35$. On the other hand, we find that the BTW data collapses 
for time and area  distributions are not compatible with the 
FSS hypothesis. The linear behavior of the moments analysis, however,
ensures that for large sizes the FSS form must be approached. 
This result can be explained if we assume that the scaling in the 
BTW model displays subdominant corrections of the
form $P(x)=(C_1 x^{-\tau_1} + C_2x^{-\tau_2} +...){\cal G}(x/x_c)$,
where $C_i$ are non-universal constants.
These corrections are compatible with the linear behavior
at large $q$, but the decay of the $P(x)$ is not
a simple power law for small $x$ and thus FSS is not obeyed.
It is worth to remark that the time and area 
distributions span over much less order of magnitude than the size 
distribution, which could explain why subdominant corrections
are more relevant in the first two cases.
Subdominant corrections 
are due to higher order operators in the dynamics and do not determine
the universality class, since the asymptotic scaling behavior is 
ruled by the leading power. 

In summary, we have presented strong numerical evidence pointing towards
a single universality class for the Manna and the BTW model.
In particular, we show that previous analyses \cite{ben,biham}
are not reliable because of systematic biases introduced by 
the method employed. Further work is needed in order to quantify 
the extent of subdominant corrections to scaling in the BTW model. 

We thank A. Barrat, D. Dhar, R. Dickman, S. Krishnamurthy,
E. Marinari, M.A. Mu\~noz 
and D. Stauffer for useful discussions and suggestions. 
The main part of the numerical simulations have been run on the {\em
Kalix} parallel computer \cite{kalix} (a Beowulf project at Cagliari
Physics Department). We thank G. Mula for leading the effort
toward organizing this computer facility.
A.V. and S.Z. acknowledge partial support 
from the European Network Contract ERBFMRXCT980183. 
The Center for Polymer Studies
is supported by NSF. A. C. acknowledge the hospitality of 
the CPS, where this work has been initiated.

\begin{narrowtext}
\begin{table}[t]
\begin{tabular}{|c|c|c|c|c|}
\hline
Model  & $\beta_s$ & $\beta_t$ & $\beta_a$ & $\tau_s$ \\
\hline \hline
Manna & $2.74\pm0.02$ & $1.50\pm0.02$ & $2.02\pm0.02$ & $ 1.27\pm0.01$\\
BTW   & $2.73\pm0.02$ & $1.52\pm0.02$ & $2.01\pm0.02$ & $ 1.27\pm0.01$\\
\hline
\end{tabular} 
\caption{Values of the critical exponents describing the scaling
of the cutoff of the distributions
for different models in $d=2$.
The results are obtained from the moments analysis (see text).
Note that the exponents $\beta_s$, $\beta_t$ and $\beta_a$ are usually 
reported in the literature as $D$, $z$ and $d_f$, respectively.}
\label{tab1}
\end{table}
\end{narrowtext}

\begin{figure}
\narrowtext
\centerline{
       \epsfxsize=7.0cm
       \epsfbox{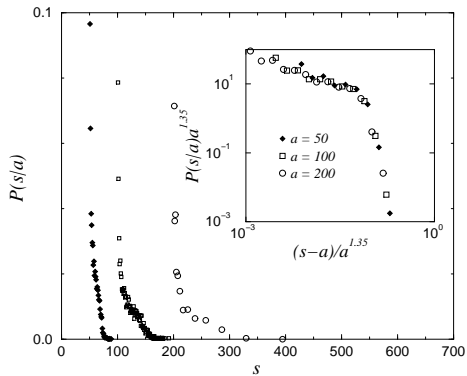}
       \vspace*{0.5cm}
        }
\caption{The figure shows the probability distribution of having an 
avalanche size $s$ given its area $a$ for the BTW model. The inset shows 
that all data collapse onto the universal scaling function 
$p(s|a)=a^{-\gamma_{sa}}f((s-a))/a^{\gamma_{sa}})$, with 
$\gamma_{sa}\simeq 1.35$.}
\label{fig1}
\end{figure}
\begin{figure}
\narrowtext
\centerline{
       \epsfxsize=7.0cm
       \epsfbox{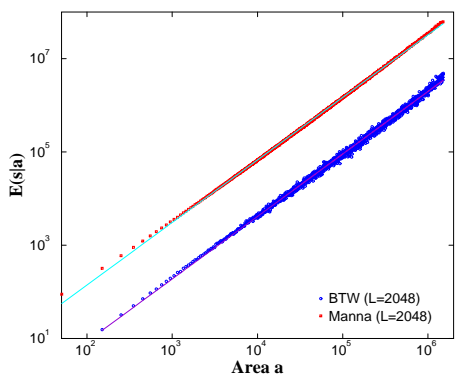}
       \vspace*{0.5cm}
        }
\caption{Conditional expectation value $E(s|a)$ for the BTW and Manna
model (after bias subtraction). The slope is given by 
$\gamma_{sa}=1.35\pm0.05$ for both curves.}
\label{fig2}
\end{figure}

\begin{figure}
\narrowtext
\centerline{
       \epsfxsize=7.0cm
       \epsfbox{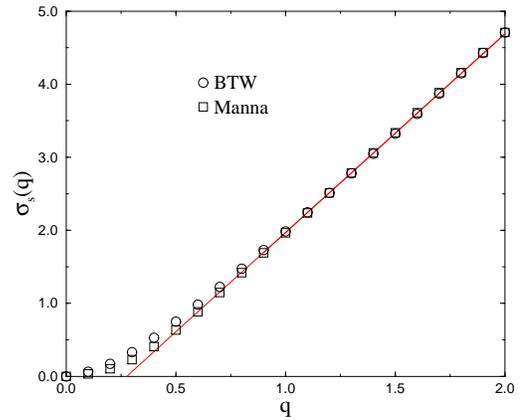}
       \vspace*{0.5cm}
        }
\caption{Plot of $\sigma_s(q)$ for the BTW and Manna model. The linear 
part has slope 2.74. Note the non universal 
corrections to the linear behavior expected
for $q \simeq \tau -1 \simeq 0.3$.}
\label{fig3}
\end{figure}

\begin{figure}
\narrowtext
\centerline{
       \epsfxsize=7.0cm
       \epsfbox{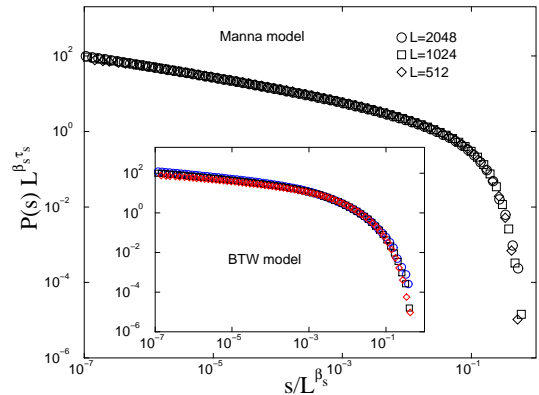}
       \vspace*{0.5cm}
        }
\caption{Data collapse analysis of the avalanche size distribution 
for the Manna and BTW (inset) models. The values used for the critical 
exponents are $\tau_s=1.27$ and $\beta_s=2.7$. Lattice  sizes used 
are reported 
in figure.} 
\label{fig4}
\end{figure}
\end{multicols}
\end{document}